\documentstyle[12pt]{article}

\def\PRL#1{{\it Phys.~Rev.~Lett.~}{\bf #1}}
\def\PRD#1{{\it Phys.~Rev.~}{\bf D#1}}

\def\NPB#1{{\it Nucl.~Phys.~}{\bf B#1}}

\def\PLB#1{{\it Phys.~Lett.~}{\bf B#1}}
\def\ARNPS#1{{\it Ann.~Rev.~Nucl.~Part.~Sci.~}{\bf #1}}
\def\ZPhysC#1{{\it Z.~Phys.~}{\bf C#1}}

\def\beq{\begin{equation}}
\def\eeq{\end{equation}}
\def\beqa{\begin{eqnarray}}
\def\eeqa{\end{eqnarray}}
\def\sm{Standard Model}
\def\tr{\tilde{t}_R}
\def\chr{\tilde{c}_R}
\def\tl{\tilde{t}_L}

\def\ur{\tilde{u}_R}
\def\ul{\tilde{u}_L}
\def\hi{\tilde h}
\def\db{\Delta\beta}

\def\cf{{\it cf.} }

\def\etal{{\it etal}}

\def\gsim{{~\raise.15em\hbox{$>$}\kern-.85em
          \lower.35em\hbox{$\sim$}~}}
\def\lsim{{~\raise.15em\hbox{$<$}\kern-.85em
          \lower.35em\hbox{$\sim$}~}}

\def\etal{{\it et al.}}

\begin{document}

\rightline{SLAC-PUB-7476}
\rightline{hep-ph/9704389}
\medskip
\rightline{April 1997}
\bigskip
\bigskip
\renewcommand{\thefootnote}{\fnsymbol{footnote}}

{\centerline{\bf SUPERSYMMETRIC BARYOGENESIS  }}
{\centerline{{\bf AND FLAVOR PHYSICS}
\footnotetext{Research supported
by the Department of Energy under contract DE-AC03-76SF00515}}}

\bigskip
{\centerline{Mihir P. Worah}}
\smallskip
\centerline {\it Stanford Linear Accelerator Center}
\centerline {\it Stanford University, Stanford, CA 94309}

\bigskip

{\centerline{\bf Abstract}}

We study the flavor physics implications of baryogenesis in the Minimal
Supersymmetric Standard Model. Enhanced
$B-\bar B$ mixing and $b \to s \gamma$ rates are generic to all
scenarios. Depending on the origin of the CP violating phase
responsible for baryogenesis there could be a large neutron electric
dipole moment, large CP violating $D-\bar D$ mixing or CP violation in
top quark production. 
We discuss how the combination of these measurements with the
requirement of baryogenesis shed light on the MSSM 
parameter space and the source of CP violation.

\newpage

In order to account for the observed baryon asymmetry of the universe,
$n_B/s=4-6 \times 10^{-11}$,
using electroweak baryon number violation one needs more effective 
CP violation and a stronger electroweak phase transition than 
present in the \sm\ \cite{ckn1}. Both of these are possible in the Minimal
Supersymmetric Standard Model (MSSM) \cite{mssm}.
Although there are large uncertainties related to the many
non-perturbative processes involved, it is still possible to obtain a
broad brush picture of the circumstances under which sufficient
baryogenesis may be acheived in the MSSM. 

There has been much recent work relating to this issue
\cite{huet-nelson,carena2,me} including the interesting possibility,
further explored in this letter, of baryogenesis in the MSSM using
just the one explicit CP violating phase present in the quark mixing
matrix \cite{me}. In this letter we present a simple, unified
description of the different mechanisms proposed above.
It is
interesting that when one puts all of it together, a coherent picture
emerges resulting in a few favored scenarios for baryogenesis in the
MSSM. Remarkably, these scenarios all have distinct experimental
predictions that can be tested at the next generation of particle
physics experiments: LHC, the $B$ factories, and improved measurements
of $d_N$, the neutron electric dipole moment (EDM). Thus, in the next few
years experiments will not only determine the possibility of 
baryogenesis in the MSSM, but also ascertain the specific mechanism
by which it occurs.

The strength of the electroweak phase transition in the MSSM could be 
enhanced due to the large coupling of a light $stop$ to the Higgs boson.
The possibility of a strong enough phase transition 
was demonstrated using a one-loop effective Higgs potential 
whenever \cite{carena1} 
\beq
m_h \lsim 80~{\rm GeV}; ~~ 
m_A \gsim 200~{\rm GeV};~~\tan\beta \lsim 2.5; ~~ 
~~m_{\tilde t_{R}} \lsim
175~{\rm GeV}; ~~ \tilde A_t \simeq 0.
\label{ewpt1}
\eeq
Here $m_h$ is the mass of the lightest (\sm-like) Higgs boson, $m_A$
is that of the pseudoscalar Higgs boson, $\tan\beta$ is the ratio of
the two Higgs vevs, 
and $\tilde A_t = A_t + \mu/\tan\beta$ is the effective $\tl -\tr$
mixing parameter. These limits are slightly relaxed if one includes
two-loop QCD effects \cite{espinosa}. 

In order to have rapid interconversion between particles and
sparticles at the phase transition, the gluinos and/or some of the
charginos are required to have masses of ${\cal O}(T_0)\sim 100$ GeV,
where $T_0$ is the critical temperature for the electroweak phase
transition. 
Besides the obvious direct search implications of these
light sparticles and Higgs boson \cite{carena2}, 
the light $\tr$ and charginos 
also result in large contributions to $B-\bar B$ mixing 
independently of the rest of the squark masses \cite{worah1}. This is
because the $b_L-\tr-\hi$ coupling proportional to the
top quark mass removes the possibility of any GIM cancellation. 

Thus we see that the requirement of a first order phase 
transition coupled to the existence of light charginos predicts 
large new contributions to $B-\bar B$ mixing.  
This new contribution to $B-\bar B$ mixing 
may be hard to detect because of the hadronic uncertainties in the \sm\
predictions. It could, however,  be resolved at the $B$ factories by 
combining the measured value of $x_d \equiv \Delta m/\Gamma$ 
with several CP violating $B$ decay asymmetries \cite{gnw}.

The most effective way to generate a cosmological particle number 
asymmetry for some species is to arrange that, during the electroweak
phase transition, a CP violating space-time dependent phase 
appears in the mass matrix for that species. If this
phase cannot be rotated away at subsequent points by the same unitary
transformation, it leads to different propagation probabilities
for particles and anti-particles, thus resulting in a particle number
asymmetry. The existence of such phases 
is possible in the MSSM if $\tan\beta$
changes as one traverses the bubble wall separating 
the symmetric phase from the broken one.\footnote{This is 
strictly true only for the leading term in an
expansion in powers of the particle mass matrices.}
Particle number asymmetries
will then be proportional to $\Delta\beta$, the change in $\beta$
across the bubble wall \cite{huet-nelson}.

It has been recently estimated that 
$\Delta\beta \propto m_h^2/m_A^2 \sim 0.01$ for the pseudoscalar Higgs
boson mass $m_A = 200-300$ GeV \cite{carena2}. 
This can actually be turned into an upper bound for $\db$ using the
relation $m_{h_+}^2=m_A^2+m_W^2$, where $m_{h_+}$ is the charged Higgs
boson mass. Charged Higgs bosons make large positive contributions
to the $b \to s \gamma$ decay rate. Although there
is a partial cancellation of this effect due to the contribution from
light charginos and $stops$ this requires fine-tuning to
be completely effective in the range of parameters considered here.
The current experimental value
for $Br(b \to s \gamma)$ already sets the limit $m_{h_+} \gsim 300$ GeV
at the $2\sigma$ level \cite{misiak}. This then implies $\db \lsim
0.01$ through the relations above.

Thus we see that the requirement of CP violation in the propagation of
particles through the bubble wall requires a light charged Higgs and
subsequently enhanced $b \to s \gamma$ decay rate \cite{huet-nelson,
carena2, me}. This scenario
will be significantly constrained at the next round of measurements at CLEO
III and at the asymmetric $B$ factories.

Baryogenesis in the MSSM proceeds most efficiently through the
generation of higgsino number or axial squark number in the bubble
wall, which then diffuses to the symmetric phase where it is processed
into baryon number.
The origin of the CP violation responsible for baryogenesis, and
consequent flavor physics effects can be understood by studying the
structure of the up-type squark mass matrix. This is justified because
for $\tan\beta \sim 1$ the effects due to the down-type squarks are
suppressed by $\sim m_b^2/m_t^2$. Further, as we will discuss below, 
the CP violation responsible for higgsino
production can be considered a special case of the ways CP
violation manifests itself in the up-type squark mass matrix and does
not lead to independent flavor physics effects.
 
Consider the mass squared matrix for the up-type squarks:
\beq
M^2_{\tilde u} = \left(\begin{array}{cc}
                 M^2_{\tilde u_{LL}} & M^2_{\tilde u_{LR}} \\
		M^{2\dagger}_{\tilde u_{LR}} & M^2_{\tilde u_{RR}}
		  \end{array} \right)
\eeq
where 
\beqa
M^2_{\tilde u_{LL}}&=&m^2_{Q}A_{U_{LL}}+(F,~D)~{\rm terms}, \nonumber \\
M^2_{\tilde u_{RR}}&=&m^2_{U}A_{U_{RR}}+(F,~D)~{\rm terms}, \nonumber \\
M^2_{\tilde u_{LR}}&=&m_Av_2\lambda_UA_{U_{LR}}+\mu v_1\lambda_U.
\eeqa
$\lambda_U$ is the Yukawa coupling matrix for up-type quarks, and the
$A_U$'s are dimensionless matrices. 
The CP violating invariant responsible for producing an asymmetry in
the right-handed up-type squark number (and hence baryon number) 
is \cite{huet-nelson}
\beq
J_{CP} = m_A|\mu |\Delta\beta 
Im Tr[e^{i\phi_B}
A^{\dagger}_{U_{LR}}\lambda^{\dagger}_U\lambda_U\rho(\ur)]
\label{jcp1}
\eeq
where $m_A$ is real (the phase information is in
$A_{U_{LR}}$), $e^{i\phi_B}$ comes from the phase of the $\mu$
parameter, and $\rho(\ur)$ can be approximated by the density matrix 
for the right-handed up-type squarks in the symmetric phase. 
%
%
Similar formulae obtain for the other squark species. 

Finally, we concentrate only on the production of $\tr$
since it is required to be light in order to enhance the
phase transition strength.
Unless they are also light, effects on the other squark species will be
Boltzmann suppressed.
For $m_{\tr}=175$ GeV, $m_{\tl}=300$ GeV, 
and $\tan\beta \sim 1$ we obtain the result
\beq
\frac{n_B}{s} \simeq
                10^{-8}\frac{\kappa\Delta\beta}{v_w}\frac{m_A}{T_0}
                \frac{|\mu |}{T_0} Im[e^{i\phi_B}
                A^{\dagger}_{U_{LR}}\lambda^{\dagger}_U\lambda_U]_{(3,3)}
\label{bnumber}
\eeq
$\kappa$ is related to the weak sphaleron rate,
$\Gamma_{ws}=\kappa \alpha_w^4 T$. There is a large uncertainty in its
precise value, with current estimates giving $\kappa = 1-0.03$
\cite{sphalerons}.
$v_w \simeq 0.1$ is the wall velocity, $\Delta\beta \lsim 0.01$,
and $T_0 \sim m_A \sim |\mu| \sim 100$ GeV is the phase transition 
temperature.
The approximations made in deriving Eq. (\ref{bnumber}) and their
validity our outlined in \cite{huet-nelson}. If $\tl$ and $\tr$ have very
different masses there is a suppression of the baryon asymmetry by
$m_{\tr}^2/m_{\tl}^2$ that is not explicit in their work. Thus the
estimate of Eq. (\ref{bnumber}) would be modified if $m_{\tl} \gg 300$
GeV. 

The CP violating phases responsible for baryogenesis 
could then logically be divided into three
separate possibilities:
\begin{itemize}

\item

There is a universal supersymmetric phase, $\phi_B$, coming from the $\mu$
term. Since the $\mu$ parameter also appears in the 
higgsino mass matrix, 
this possibility also results in the production of 
higgsino number, which contributes
to the baryogenesis an amount similar to that of the axial $stop$
number when the higgsino mass parameters are all $\sim T_0 \simeq 100$
GeV \cite{huet-nelson}.%
\footnote{This possibility was recently studied in detail
\cite{carena2}. They conclude that baryogenesis 
from higgsino production is possible only for very specific 
choices of the mass parameters, and that it is not possible 
at all from axial $stop$ production. We feel this conclusion is too
strong given the inherent uncertainties in baryogenesis calculations.} 

\item

There are flavor dependent supersymmetric phases present in
$A_{U_{LR}}$. Note that a universal phase in $A_{U_{LR}}$ 
can be rotated into the higgsino mass, and so is not
distinct from the previous scenario. 

\item

The supersymmetric parameters are all real, and the only phases are in
the quark mass matrix, $\lambda_U$. This allows the possibility that
there is only one large fundamental phase, that of the CKM matrix,
that is reponsible for both the baryon asymmetry and the CP
violating $K-\bar K$ mixing \cite{me}.%

\end{itemize}

We will now consider these three possibilities separately. 
One should realise, however, that in the most general case 
phases from all three sources could contribute.

The presence of a phase, $\phi_B$, for the $\mu$ term leads
to a neutron EDM. The experimental bound $d_N \le 1 \times
10^{-25}$ e-cm \cite{pdg} tells us that either 
$\phi_B \lsim 10^{-2}$ or $m_{\tilde u} \gsim 1$ TeV \cite{edm}
where $m_{\tilde u}$ is the average first generation squark mass. 
Using a diagonal and real $A_{U_{LR}}$,
and top quark Yukawa coupling $\lambda_t = 1$, one obtains from
Eq. (\ref{bnumber}) the requirement $\phi_B \gsim 10^{-2}$ for $\kappa
=1,~\db=0.01$. 
Thus, either the neutron EDM will be discovered soon or 
the first generation squarks are heavy \cite{carena2}.
Any reduction in the values of some of the parameters used to evaluate 
Eq. (\ref{bnumber}) would force $\phi_B$ to be larger, 
and hence the first generation
squarks to be heavier. In particular if, $\kappa \ll 1$ or $\db \ll
0.01$, one would require $\phi_B \sim 1$ to get a large enough baryon
asymmetry. The constraint coming from $d_N$ would then lead us to the
particular realization of supersymmetry known as Effective
Supersymmetry \cite{eff-susy1}, 
where the first (two) generations of squarks have masses
larger than 1 TeV while the third generation is light. 
Some of the flavor physics implications of this 
model have been studied in \cite{eff-susy2, yuval-me}. 

If $\phi_B \sim 1$, 
an additional signal of its presence in the $\tl-\tr$
mixing would be large CP violating asymmetries in $t\bar t$ production at 
hadron colliders. This manifests itself as an asymmetry
in the transverse energy distribution of the lepton and antilepton
decay products of the $t\bar t$ pair which could be
large enough to observe at the LHC \cite{schmidt}.

The possibility that the phase arises in $A_{U_{LR}}$ allows us to
evade the constraint from the neutron EDM because in this case 
the phase in the
$\tl-\tr$ mixing is independent of that of $\ul-\ur$ mixing.
Using $\kappa=1,~\db=0.01$, and a diagonal  $A_{U_{LR}}$ in
Eq. (\ref{bnumber}), 
the requirement of baryogenesis implies 
$\phi_t \gsim 10^{-2}$ for the phase of $A_t \equiv A_{U_{LR}}(3,3)$. 
If however $\kappa \ll 1$ or $\db \ll 0.01$, we would require
$\phi_t \sim 1$, leading to the possibility of 
CP violating $t\bar t$ production mentioned above. 
The absence of a large neutron EDM even in the presence of light
first generation squarks would distinguish this scenario from the
previous one. 
The supersymmetry breaking scale in this case cannot be
too high, else one would generate too large a neutron EDM 
due to RGE effects \cite{jim}. 

Finally we come to the third possibility that the supersymmetric
parameters $A_{U_{LR}}$ and $\mu$ are real, with all the CP violation
being in the quark mass matrix \cite{me}. Notice that
$\lambda^{\dagger}_U\lambda_U$ in Eq. (\ref{bnumber}) is Hermitian,
hence the phase is on one of the off-diagonal terms. 
One then requires $A_{U_{LR}}$ to have off diagonal entries in order
to move this phase to the (3,3) element of the product
$A^{\dagger}_{U_{LR}}\lambda^{\dagger}_U\lambda_U$.%
\footnote{A scenario where the lightest squark is an admixture of
$\chr$ and $\tr$ was considered in \cite{me} as a way to 
motivate large off-diagonal entries in $A_{U_{LR}}$.
The predictions for low energy flavor physics in
that scenario are not very different from those obtained here.}
Given the reasonable assumption that at least part of $\theta_C$,
the Cabbibo angle, is generated in the up-type quark mass matrix 
\cite{nir-seiberg}, off-diagonal terms in $A_{U_{LR}}$ 
always lead to large $D-\bar D$ mixing due to gluino mediated box 
diagrams. The magnitude of the mixing is generically within an order
of magnitude of the current experimental bound $\Delta(m_D) < 1.3
\times 10^{-13}$ GeV \cite{pdg}. Further, given the hierarchical
structure of the quark masses and mixings, 
one expects the largest off-diagonal entry in 
$\lambda^{\dagger}_U\lambda_U$ to be $\sim \theta_C^2 \sim 0.04$.
For example the ansatz $\lambda_U = V^{\dagger}_{CKM}\hat
{\lambda}_U V_{CKM}$ where $V_{CKM}$ is the CKM matrix, and $\hat
\lambda_U$ is the diagonal matrix of up-type Yukawa couplings can lead to
\beq
Im[A^{\dagger}_{U_{LR}}\lambda^{\dagger}_U\lambda_U]_{(3,3)} =
\lambda_t^2|V_{cb}|\sin\gamma 
\eeq
for
\beq
A_{U_{LR}}=\left(\begin{array}{ccc}
            1&0&0 \\
            0&1&1 \\
            0&1&1 \end{array}\right),
\eeq
where $\gamma \sim 1$ is the phase in the CKM matrix, that is measured in
$K-\bar K$ mixing. This
leads to a large enough baryon asymmetry [\cf Eq. (\ref{bnumber})] for
$\kappa = 1,~\db=0.01$, but is ruled out if $\kappa \ll 1$ or $\db
\ll 0.01$. Thus we see that the three different ways in which the CP
violation required for baryogenesis 
can manifest itself in the up-type squark mass matrix all result in
experimentally distinguishable scenarios. 

To summarize the effects on flavor physics imposed
by the requirement of electroweak baryogenesis in the MSSM:
 
A sufficiently first order phase transition requires a light
Higgs boson and $\tr$. The light $\tr$ coupled with the existence 
of light charginos required to
convert sfermions to fermions implies large new contributions to
$B-\bar B$ mixing. These could be observed at the asymmetric $B$
factories. 

In order to get a CP violating asymmetry in the
propagation of particles through the bubble wall, one requires a
non-trivial variation in the ratio of the Higgs vevs. This implies
that the second Higgs doublet cannot be much heavier than the first.
This scenario will be significantly constrained by improvements in the
experimental accuracy for the $b\to s\gamma$ decay rate.

The CP violating phase responsible for baryogenesis 
resides in the up-type squark mass matrix. It could be a universal
supersymmetric phase in which case either the neutron EDM will be
discovered soon, or the first generation squarks are heavy. This
scenario will be most significantly tested by 
improvements in the measurement of $d_N$ combined with 
direct searches for first generation squarks. 

The CP violation could also come from a 
flavor dependent phase in $A_{U_{LR}}$.
This scenario could be distinguished from the one above if light first
generation squarks were discovered, but not the
neutron EDM. 

Alternatively the
supersymmetric parameters could be real, and the phase could come from
the quark mixing matrix. This scenario predicts $D-\bar D$ mixing 
at a level that should be discovered soon.
This scenario is the most constrained of the three, since the size of
the CP violating invariant can be estimated from our knowledge of the
quark masses and mixings, and is suppressed by the small angle 
$V_{cb}$. 

Thus we see that the possibility of baryogenesis in the MSSM
significantly constrains its parameter space. Experiments planned for
the next few years will shed light on this picture. 
Direct searches for the light particles required are the first step
towards determining the possibility of baryogenesis in the MSSM. If
these particles are discovered, there are currently three  
different scenarios for the source of the CP violating phase 
that allow baryogenesis, all having distinct and testable experimental
consequences. 
The flavor physics effects discussed here will then serve to elucidate 
the mechanism for baryogenesis in the MSSM.

\bigskip

Useful discussions with A. Grant, Y. Grossman, P. Huet, M. Peskin, and
J. Wells are happily acknowledged.


\begin{thebibliography}{99}
\bibitem{mssm}
{By Minimal Supersymmetric Standard Model we mean the supersymmetric
extension of the standard model with minimal particle content and $R$
parity conservation. The
constrained MSSM includes the assumptions of coupling constant
unification, universal scalar and gaugino masses and 
universal trilinear terms at a high
``unification'' scale.}
\end{thebibliography}
\end{document}